# On Abreu's theory of space and time and new measurements of absolute velocities

Gustavo Homem
Graduate student at I.S.T , Universidade Técnica de Lisboa
Exchange student at NBIfAFG , University of Copenhagen
email: ghomem@ist.utl.pt

**Abstract**

Abreu, R. suggested a new interpretation of the Lorentz Transformation (LT) [3], a different LT derivation [2] and a new "synchronized" space−time transformation (ST) [4]. While he has shown that his results could be expressed in a Lorentz−invariant form, the simplest (but perhaps most shocking) approach appears to be the use of the ST. The simplicity of this formulation, which bypasses the well known paradoxes from Relativity, along with the fact that phenomena like time dilatation are still predicted, suggests the need for further investigation on this subject. We briefly discuss this new space−time approach, using ST, and suggest two new methods for measuring absolute velocities which also provide a way to test Special Relativity.

# 1. Introduction

One of the problems Relativity intended to solve was the definition of a common time, with one clock per point, for every frame of reference. Without this, it is not possible to measure velocities at all, since the Newtonian way of measuring time, with a single observer, does not yield unique time intervals for a given pair of events[1][1]. Einstein's synchronization method was suggested for this purpose as he postulated the speed of light to be the same in every frame of reference. On the other hand, based on the relation between rhythms of clocks in different frames of reference S and S',

$$\Delta t' = \Delta t \sqrt{1 - \frac{v^2}{c^2}} \qquad (1)$$

Abreu concluded that not to be possible [2]. This relation between rhythms of two clocks, one on S and the other on S', seems to hold regardless of how the readings on adjacent clocks compare, on each of these frames. It is, in fact, important to be aware of the difference between time intervals and clock readings [2,3].

At first, Einstein calls S the "stationary frame" while S' is called the "moving frame" [1]. We will follow this naming convention. Once Abreu finds the speed of light for a generic moving frame he suggests the use of Einstein's method of synchronization [4], with the correct speed value, to synchronize the clocks of that moving frame.

In short, from the following assumptions:

1) There is one frame of reference for which the speed of light emitted from a standing source is c in every direction.
2) In the frame of reference where 1) holds the speed of light is independent of the state of motion of the source[2]
3) The average two-way light speed is c on every frame of reference[3]

and not making use of the principle of relativity, a transformation of coordinates can be obtained in the form,

$$t' = t \sqrt{1 - \frac{v^2}{c^2}} \qquad (2)$$

$$x' = \frac{x - vt}{\sqrt{1 - \frac{v^2}{c^2}}} \qquad (3)$$

A common time is defined for every frame of reference, but it passes at different rhythms. Within each frame the clocks are synchronized (t' does not depend on x or x') but they are not synchronized with the ones in another frame as the common times differ.

It is easy to see that the well known twins paradox takes a different form with this transformation. By imposing fixed lengths and times on one frame and comparing with the ones, yielded by (2)–(3), for the other it can be seen that there are still disagreements between twins (and it had to be so, since the time and length scales depend on the frame of reference). However, while each one considers himself to be "normal", they do make opposite statements about each other (unlike with Relativity where they state the same thing) and that could hardly be called a paradox.

The speed of light in a generic frame is found to have two possible values, depending on the direction of propagation

---

1  The measured time interval would depend on the position of the observer.
2  This is usually expressed as part of 1) but we put it separately to stress that once 1) is accepted 2) can easily be put to test.
3  This is well a know experimental fact. "c" is the average two-way speed although being often called "the speed of light".

$$c_v^+ = \frac{c-v}{1-\frac{v^2}{c^2}} \tag{4}$$

$$c_v^- = \frac{c+v}{1-\frac{v^2}{c^2}} \tag{5}$$

An object that moves according to x=ut on S, will have a relative velocity on S' of:

$$u_v = \frac{u-v}{1-\frac{v^2}{c^2}} \tag{6}$$

where $u_v$ stands for the velocity of an object that moves with absolute velocity u, relative to a frame with absolute velocity v. The reciprocal relation

$$v_u = \frac{v-u}{1-\frac{u^2}{c^2}} \tag{7}$$

shows that the relative velocities are not symmetrical unless the absolute velocities are symmetrical themselves. This is due to different clock rhythms, which depending on the square of the absolute velocities, can only be equal in the case of equal or symmetrical absolute velocities.

According to (2) and (3), the generic space–time transformation between two arbitrary moving frames S' and S'' is found to be

$$t'' = t' \frac{\sqrt{1-\frac{u^2}{c^2}}}{\sqrt{1-\frac{v^2}{c^2}}} \tag{8}$$

$$x'' = x' \frac{\sqrt{1-\frac{v^2}{c^2}}}{\sqrt{1-\frac{u^2}{c^2}}} + t' \frac{(v-u)}{\sqrt{1-\frac{v^2}{c^2}}\sqrt{1-\frac{u^2}{c^2}}} \tag{9}$$

A generic transformation for relative velocities can also be easily derived, from which is easily seen that the speed of light is independent of the motion of the source. That is, in this context, the speed of light depends on the frame of observation but not on the motion of the source relative to that frame.

The apparent problem of this approach is that the absolute velocities are unknown and so it may be said not to be possible to use Einstein's method of synchronization on a generic moving frame, unless we postulate that the speed of light is c. This is usually accepted as a justification for stating that all frames may be regarded as stationary, but as was pointed out before [4], not knowing which one is the stationary frame differs from stating that all frames are stationary, being the latter statement one possible way of expressing the principle of Relativity.

Under the assumptions of this theory, it is possible to obtain absolute speeds from simple experiments and therefore finding the stationary frame. One possible way was already proposed [4]. We will suggest two other methods of greater simplicity.

## 2. A procedure for measuring absolute speeds

Let's consider two moving frames, S' and S'', whose (yet unknown) absolute velocities are respectively v and u. There are two mirrors on S'' standing on positions x' and −x'. An observer and a clock are standing in the origin of S'. When t'=0, t''=0 and the origins of S' and S'' overlap. By that time, two rays of light are emitted from the origin of S'' towards the two mirrors on S'. After reaching the mirror each ray is reflected back to the origin of S''.

Let's consider the trip of the ray emitted on the positive direction[4]. Its motion is described in S'' by

$$x'' = c_u^+ t'' \tag{10}$$

On the other hand, the motion of the mirror on the positive side is given by

$$x'' = x_0'' + v_r t'' \tag{11}$$

where x''$_0$ is the coordinate, on S'', corresponding to x' when t'=t''=0. Thus, the time for the first part of the journey is

$$\Delta t''_1 = \frac{x_0''}{c_u^+ - v_r} \tag{12}$$

where v$_r$ is the velocity of S' relative to S''.

The second part of the journey, as seen from S'', is towards a fixed target: its own origin. However, the path length depends on how far, from the origin of S'', the mirror on S' got until it was reached by the ray of light. This means

$$\Delta t''_2 = \frac{x_0'' + v_r \Delta t_1''}{c_u^-} \tag{13}$$

or

$$\Delta t''_2 = \frac{x_0''}{c_u^-}\left(1 + \frac{v_r}{c_u^+ - v_r}\right) \tag{14}$$

Thus, the total travel time is

$$\Delta t'' = \frac{x_0''(c_u^- + c_u^+)}{c_u^-(c_u^+ - v_r)} \tag{15}$$

From (9) we have

---

[4] The assignment of c$^+_u$ to one particular direction defines the orientation of the observer's frame and consequently of all the others, since the theory was developed with co–oriented frames. The values of u and v (namely their signs) obtained from the experiment should be interpreted according to this orientation.

$$x_0'' = x' \frac{\sqrt{1-\frac{v^2}{c^2}}}{\sqrt{1-\frac{u^2}{c^2}}} \tag{16}$$

which along with the appropriate replacements conforming to (4),(5) and (7) yields

$$\Delta t''_+ = \frac{2c\,x'\sqrt{1-\frac{v^2}{c^2}}\sqrt{1-\frac{u^2}{c^2}}}{(c-v)(c-u)} \tag{17}$$

In a similar manner, the total time for the light ray emitted in the negative direction is found to be

$$\Delta t''_- = \frac{2c\,x'\sqrt{1-\frac{v^2}{c^2}}\sqrt{1-\frac{u^2}{c^2}}}{(c-u)(c+v-2u)} \tag{18}$$

Equations (17) and (18) can be compared with the experimental values and solved for u and v.

## 3. Absolute speed measurements using the Doppler effect

Consider the same *setup* used for the previous experiment, except that the mirrors are replaced by light sources of characteristic period $T_0$. The relation between time on the two frames is given by (8). However, the measurement of, say, the period of a light source, on a particular frame of reference can only be done by multiple observers with synchronized clocks. The period of a moving source, as seen by a single standing observer, is affected by delays, implied by the finiteness of the speed of light. This apparent alteration of the period is commonly known, through its associated frequency shift, as the *Doppler effect*. This effect is, in fact, far more general than a frequency shift. Indeed, it is a question of how time apparently passes on a point that is moving relative to the observer. [5]

Let's suppose that the origin of S' is moving away, from the observer in the origin of S'', in the positive direction and with a relative velocity $v_r$. An event which takes place there on time $t'_0$ is seen by the observer in the instant

$$t_{obs} = t'_0 \frac{\sqrt{1-\frac{u^2}{c^2}}}{\sqrt{1-\frac{v^2}{c^2}}} + \frac{v_r}{c_u^-} t''_0 \tag{19}$$

where $v_r$ is positive and light propagates in the negative direction (see also note 4). From (8), (19) can be written

---

[5] The existence of this phenomenon suggests that care should be taken in the usage of expressions, such as, "the observer sees" which are often intended to mean "a set of observers measure"

$$t_{obs} = t'_0 \left(1 + \frac{v_r}{c_u^-}\right) \frac{\sqrt{1 - \frac{u^2}{c^2}}}{\sqrt{1 - \frac{v^2}{c^2}}} \qquad (20)$$

This expression holds for an arbitrary instant. Thus, writing (20) for two arbitrary instants and subtracting we find

$$\Delta t_{obs} = \Delta t'_0 \left(1 + \frac{v_r}{c_u^-}\right) \frac{\sqrt{1 - \frac{u^2}{c^2}}}{\sqrt{1 - \frac{v^2}{c^2}}} \qquad (21)$$

If S' is moving away in the negative direction, a similar formula is obtained where $c_u^+$ is used instead of $c_u^-$.

This reasoning can be generalized for any point on S', moving away or towards the observer, being the result

$$\Delta t_{obs} = \Delta t'_0 \left(1 \pm \frac{v_r}{c_u^{+-}}\right) \frac{\sqrt{1 - \frac{u^2}{c^2}}}{\sqrt{1 - \frac{v^2}{c^2}}} \qquad (22)$$

where $v_r$ is the speed of the moving point relative to the observer. The "−" sign should be picked if the motion is towards the observer.

These results can be directly applied to the period of a moving source. Since, on the frame of the moving source, its period is the characteristic period $T_0$, we have

$$T_{obs} = T_0 \left(1 \pm \frac{v_r}{c_u^{+-}}\right) \frac{\sqrt{1 - \frac{u^2}{c^2}}}{\sqrt{1 - \frac{v^2}{c^2}}} \qquad (23)$$

Now let's consider two sources on fixed points of a frame of reference S', as explained before. From (4), (5) and (7) we get

$$T_{obs} = T_0 \left(1 + \frac{v - u}{c + u}\right) \frac{\sqrt{1 - \frac{u^2}{c^2}}}{\sqrt{1 - \frac{v^2}{c^2}}} \qquad (24)$$

for the moving away source and

$$T_{obs} = T_0 \left(1 - \frac{v-u}{c-u}\right) \frac{\sqrt{1-\frac{u^2}{c^2}}}{\sqrt{1-\frac{v^2}{c^2}}} \tag{25}$$

for the other one. Thus, the frequency shifts are, respectively:

$$f_{obs} = f_0 \left(\frac{c+u}{c+v}\right) \frac{\sqrt{1-\frac{v^2}{c^2}}}{\sqrt{1-\frac{u^2}{c^2}}} \tag{26}$$

$$f_{obs} = f_0 \left(\frac{c-u}{c-v}\right) \frac{\sqrt{1-\frac{v^2}{c^2}}}{\sqrt{1-\frac{u^2}{c^2}}} \tag{27}$$

Equations (26) and (27) can be compared with the experimental values and solved for u and v.

It is worth noting that when u=0, *i.e.*, when the observer is in the stationary frame, (26) and (27) reduce to:

$$f_{obs} = f_0 \frac{c}{c \pm v} \sqrt{1-\frac{v^2}{c^2}} \tag{28}$$

and that when v<<c the classical result is found:

$$f_{obs} = f_0 \frac{c}{c \pm v} \tag{29}$$

## 4. Practical considerations

These two experiments, which were presented in a rather abstract way, are actually quite simple. All that is necessary is a moving rod with two mirrors, or two light sources, in its ends, and a way to make sure that the rod moves with constant speed, relative to the sensor which performs the measurements and is located between its two ends. Obviously, since we are interested in measuring a vector and not only one component there should be one *setup* for each direction.

Now, the whole point of these experiments is finding a way to synchronize clocks in order to correctly measure speeds. While an Einstein–like synchronization procedure, with an arbitrary *a priori* value for the speed of light, might not yield the correct values for the speed of an object, it still enables us to discover whether or not its speed is constant. Thus, it seems possible to ensure that the moving rods used in the experiments have a constant relative velocity. Meanwhile, we cannot guarantee that the observer's absolute velocity is not changing as we assume throughout the text. In fact, for an observer in the Earth it is expected to change. Nevertheless, if it changes slowly when compared to the times involved in the movement of the rods (which can most likely be made very small) the experiments should yield useful results and could even be used to investigate the time evolution of the absolute velocity of the Earth.

## 5. Final remarks

The same experiments here described can also be analyzed in the context of Special Relativity. It can be shown that for the first one we would obtain equations (17) and (18) with the u parameter set to zero. For the second experiment equations (28) would be found. This is not surprising since in Special Relativity every "inertial" frame is said to be stationary. Thus, these experiments provide a straight forward way of comparing both theories. A time varying value for u would support Abreu's view while a constant null result would corroborate Relativity.